\documentclass[twocolumn,usenames,dvipsnames]{aastex61}

\usepackage{graphicx}
\usepackage{float}
\usepackage{amsmath}
\usepackage{acronym}
\usepackage{multirow}
\usepackage{ifthen}
\usepackage{blindtext}
\usepackage{svn-multi}
\usepackage[normalem]{ulem}
\usepackage{hyperref}
\usepackage{etoolbox}
\usepackage{fancyhdr}
\usepackage{import}
\def\apjl{Astrophysical Journal}             % Astrophysical Journal, Letters

\def\apjs{ApJS}              % Astrophysical Journal, Supplement
\def\nat{Nature}              % Nature
\def\aap{A\&A}                % Astronomy and Astrophysics

\DeclareGraphicsExtensions{%
    .pdf,.PDF,%
    .png,.PNG,%
    .jpg,.mps,.jpeg,.jbig2,.jb2,.JPG,.JPEG,.JBIG2,.JB2,%
}

\renewcommand{\today}{\number\day\space\ifcase\month\or
  January\or February\or March\or April\or May\or June\or
  July\or August\or September\or October\or November\or December\fi
  \space\number\year}

\newcommand{\macro}[1]{\textcolor{black}{#1}}

\newcommand{\Msun}{\ensuremath{{M}_\odot}}

\newcommand{\EventName}{\macro{GW170608}}
\newcommand{\EventTime}{\macro{02:01:16.49 UTC}}
\newcommand{\EventDate}{\macro{June 8, 2017}}
\newcommand{\SpecialAnalysisFLower}{\macro{30\,Hz}}

\newcommand{\SpecialAnalysisCoincTime}{\macro{1.2\;days}}
\newcommand{\SpecialAnalysisStart}{\macro{June 7, 2017}}
\newcommand{\SpecialAnalysisEnd}{\macro{June 9, 2017}}
\newcommand{\EventSourceFrameMTOTNoErrors}{\macro{19}}
\newcommand{\EventNetworkSNR}{\macro{13}}

\newcommand{\EventFARPyCBC}{\macro{1 in 3,000 years}}  % NB upper bound!!
\newcommand{\EventFARGstLALSpecialAnalysis}{\macro{1 in 160,000 years}}
\newcommand{\EventFARcWBUnClean}{\macro{1 in ${\sim} 30$ years}}

\newcommand{\EventHLAbsTimeDelay}{\ensuremath{7\,\mathrm{ms}}}
\newcommand{\EventSkyAreaBayestar}{\macro{\ensuremath{860\,\mathrm{deg}^2}}}

\newcommand{\MTOTSCOMPACTBoxing}{\macro{\ensuremath{21.8_{-1.7}^{+5.9}}}} % PE: Overall source (physical) total mass
\newcommand{\MCSCOMPACTBoxing}{\macro{\ensuremath{8.9_{-0.3}^{+0.3}}}} % PE: Overall source (physical) chirp mass

\newcommand{\rateintrbracket}{\ensuremath{\macro{12\text{--}213}~\mathrm{Gpc^{-3}\,yr}^{-1}}}

\newcommand{\MTOTSCOMPACTJune}{\macro{\ensuremath{19_{-1}^{+5}}}} % PE: Overall source (physical) total mass
\newcommand{\MCSCOMPACTJune}{\macro{\ensuremath{7.9_{-0.2}^{+0.2}}}} % PE: Overall source (physical) chirp mass
\newcommand{\MONESCOMPACTJune}{\macro{\ensuremath{12_{-2}^{+7}}}} % PE:Overall source (physical) mass m1
\newcommand{\MTWOSCOMPACTJune}{\macro{\ensuremath{7_{-2}^{+2}}}} % PE: Overall source (physical) mass m2
\newcommand{\MASSRATIOCOMPACTJune}{\macro{\ensuremath{0.6_{-0.4}^{+0.3}}}} % PE: Overall mass ratio q
\newcommand{\CHIEFFCOMPACTJune}{\macro{\ensuremath{0.07_{-0.09}^{+0.23}}}} % PE: Overall effective spin parameter chi_eff
\newcommand{\DISTANCECOMPACTJune}{\macro{\ensuremath{340_{-140}^{+140}}}} % PE: Overall distance in Mpc
\newcommand{\REDSHIFTCOMPACTJune}{\macro{\ensuremath{0.07_{-0.03}^{+0.03}}}} % PE: Overall redshift

\newcommand{\SPINFINALinplaneCOMPACTJune}{\macro{\ensuremath{0.69_{-0.05}^{+0.04}}}} % PE: Overall final spin af with in-plane components
\newcommand{\MFINALSavgCOMPACTJune}{\macro{\ensuremath{18.0_{-0.9}^{+4.8}}}} % PE: Overall source final mass mf with updated formula

\newcommand{\ERADCOMPACTJune}{\macro{\ensuremath{0.85_{-0.17}^{+0.07}}}} % PE: Overall radiated energy in solar masses
\newcommand{\LPEAKCOMPACTJune}{\macro{\ensuremath{3.4_{-1.6}^{+0.5} \times 10^{56}}}} % PE: Overall peak luminosity in erg/s
\newcommand{\PESKYNINTYJune}{\macro{\ensuremath{520~\mathrm{deg^2}}}} % LALInference: 90% credible region

\acrodef{BBH}[BBH]{binary black hole}
\acrodef{SNR}[SNR]{signal-to-noise ratio}
\acrodef{LIGO}[LIGO]{Laser Interferometer Gravitational-Wave Observatory}
\acrodef{LHO}[LHO]{The LIGO Hanford Observatory}
\acrodef{LLO}[LLO]{the LIGO Livingston Observatory}
\acrodef{LSC}[LSC]{LIGO Scientific Collaboration}
\acrodef{O1}[O1]{first observing run}
\acrodef{O2}[O2]{second observing run}
\acrodef{GW}[GW]{gravit\-ational wave}
\acrodef{BH}[BH]{black hole}
\acrodef{GR}[GR]{general relativity}
\acrodef{NR}[NR]{numerical relativity}
\acrodef{PN}[PN]{post-Newtonian}
\acrodef{PSD}[PSD]{power spectral density}
\acrodef{LMXBs}[LMXBs]{Low-mass X-ray binaries}
\acrodef{LMXB}[LMXB]{low-mass X-ray binary}

\newcommand{\GW}[0]{\ac{GW}\xspace}
\newcommand{\SNR}[0]{\ac{SNR}\xspace}
\newcommand{\LIGO}[0]{\ac{LIGO}\xspace}

\newcommand{\GR}[0]{\ac{GR}\xspace}

\begin{document}

\title{\EventName: Observation of a \EventSourceFrameMTOTNoErrors-solar-mass binary black hole coalescence}

\def\allauthors{%% this conditional keeps \allauthors from turning on
%%                 unless \AuthorCollaborationLimit is used:
%%  DT: But in fact it does not work because \AuthorCollaborationLimit 
%%  is set to 100 in the cls file.
\ifnum\AuthorCollaborationLimit>0
 \ifnumlines\nolinenumbers\fi
 \onecolumngrid
 \AuthorCollaborationLimit=3000  %%AASTeX is 100
 \largestAffilNum=300            %%AASTeX is 100
{
\fi
}
}

\allauthors

\author{B.~P.~Abbott}
\affiliation{LIGO, California Institute of Technology, Pasadena, CA 91125, USA}
\author{R.~Abbott}
\affiliation{LIGO, California Institute of Technology, Pasadena, CA 91125, USA}
\author{T.~D.~Abbott}
\affiliation{Louisiana State University, Baton Rouge, LA 70803, USA}
\author{F.~Acernese}
\affiliation{Universit\`a di Salerno, Fisciano, I-84084 Salerno, Italy}
\affiliation{INFN, Sezione di Napoli, Complesso Universitario di Monte S.Angelo, I-80126 Napoli, Italy}
\author{K.~Ackley}
\affiliation{University of Florida, Gainesville, FL 32611, USA}
\affiliation{OzGrav, School of Physics \& Astronomy, Monash University, Clayton 3800, Victoria, Australia}
\author{C.~Adams}
\affiliation{LIGO Livingston Observatory, Livingston, LA 70754, USA}
\author{T.~Adams}
\affiliation{Laboratoire d'Annecy-le-Vieux de Physique des Particules (LAPP), Universit\'e Savoie Mont Blanc, CNRS/IN2P3, F-74941 Annecy, France}
\author{P.~Addesso}
\affiliation{University of Sannio at Benevento, I-82100 Benevento, Italy and INFN, Sezione di Napoli, I-80100 Napoli, Italy}
\author{R.~X.~Adhikari}
\affiliation{LIGO, California Institute of Technology, Pasadena, CA 91125, USA}
\author{V.~B.~Adya}
\affiliation{Max Planck Institute for Gravitational Physics (Albert Einstein Institute), D-30167 Hannover, Germany}
% ...
\collaboration{LIGO Scientific Collaboration and Virgo Collaboration}

\begin{abstract}
On \EventDate{} at \EventTime{}, a gravitational-wave signal from the merger of two stellar-mass black holes was observed by the two Advanced LIGO detectors with a network signal-to-noise ratio of~\EventNetworkSNR{}. 
This system is the lightest black hole binary so far observed, with component masses $\MONESCOMPACTJune\,\Msun$ and $\MTWOSCOMPACTJune\,\Msun$ (90\% credible intervals).  These lie in the range of measured black hole masses in low-mass X-ray binaries, thus allowing us to compare black holes detected through gravitational waves with electromagnetic observations. 
The source's luminosity distance is $\DISTANCECOMPACTJune~\mathrm{Mpc}$, corresponding to redshift $\REDSHIFTCOMPACTJune$.  We verify that the signal waveform is consistent with the predictions of general relativity.
\end{abstract}

\section{Introduction}
\label{s:intro}

The first detections of binary black hole mergers were made by the Advanced
\LIGO \citep{TheLIGOScientific:2014jea,GW150914-DETECTORS} during its \ac{O1} in 
2015 \citep{GW150914-DETECTION,GW151226-DETECTION,O1:BBH}.
Following a commissioning break, LIGO undertook a \ac{O2} from 
November 30, 2016 to August 25, 2017, with the Advanced Virgo
detector \citep{TheVirgo:2014hva} joining the run on August 1, 2017.  Two
binary black hole mergers \citep{GW170104-DETECTION,GW170814-DETECTION} and one
binary neutron star merger \citep{GW170817-DETECTION} have been reported in
\ac{O2} data.  Here we describe \EventName{}, a binary black
hole merger with likely the lowest mass of any so far observed by \LIGO. 

\EventName{} was first identified in data from \ac{LLO}, which was in normal observing mode.  \ac{LHO} was operating stably with sensitivity typical for \ac{O2}, but its data were not analyzed automatically as the detector was undergoing a routine angular control procedure (Section~\ref{s:detectors} and Appendix~\ref{s:app_a2l}). 
Matched-filter analysis of a segment of data around this time revealed a candidate with source parameters consistent between both \ac{LIGO} detectors; further offline analyses of a longer period of data confirmed the presence of a \GW signal from the coalescence of a binary black hole system, with high statistical significance (Section~\ref{s:observation}).

The source's parameters 
were estimated via coherent Bayesian analysis \citep{Veitch:2014wba,GW150914-PARAMESTIM}. 
A degeneracy between the component masses $m_1$, $m_2$ prevents precise determination of their individual values, but the chirp mass $\mathcal{M} = (m_1m_2)^{3/5}(m_1+m_2)^{-1/5}$ is well measured and is the smallest so far observed for a merging black hole binary system, with the total mass $M = m_1 + m_2$ also likely the lowest so far observed (Section~\ref{s:parameters}). 
Individual black hole spins are poorly constrained, however we find a slight preference for a small positive net component of spin in the direction of the binary orbital angular momentum.

In combination with GW151226 \citep{GW151226-DETECTION}, this system points to a population of black hole binaries with component masses comparable to those of black holes found in X-ray binaries (Section~\ref{s:astro}) and significantly below those seen in other LIGO-Virgo black hole binaries. 

We also test the consistency of the observed \GW signal with the predictions of \GR; we find no deviations from those predictions.

\section{Detector operation}
\label{s:detectors}

The \LIGO detectors measure gravitational-wave strain using dual-recycled Michelson interferometers with Fabry-Perot arm cavities \citep{TheLIGOScientific:2014jea,GW150914-DETECTORS}.
During O2, the horizon distance for systems with component masses similar to \EventName{}---the distance at which a binary merger optimally oriented with respect to a detector has an expected \ac{SNR} of 8 \citep{Allen:2005fk,Chen:2017wpg}---peaked at ${\sim} 1$\,Gpc for \ac{LLO}, and at ${\sim} 750$\,Mpc for \ac{LHO}.

At the time of \EventName{}, \ac{LLO} was observing with a sensitivity close to its peak.
\ac{LHO} was operating in a stable configuration with a sensitivity of ${\sim}650$\,Mpc; a routine procedure to minimize angular noise coupling to the strain measurement was being performed \citep{A2LBeamPos}.
Although such times are in general not included in searches, it was determined that \ac{LHO} strain data were unaffected by the procedure at frequencies above \SpecialAnalysisFLower{}, and may thus be used to identify a GW source and measure its properties.  More details on \ac{LHO} data are given in Appendix~\ref{s:app_a2l}.

Similar procedures to those used in verifying previous \GW detections \citep{GW170814-DETECTION} were followed and indicate that no disturbance registered by LIGO instrumental or environmental sensors \citep{Effler:2014zpa} was strong enough to have caused the \EventName{} signal.

Calibration of the \ac{LIGO} detectors is performed by inducing test-mass motion using photon pressure from modulated auxiliary lasers \citep{Karki:2016pht, GW150914-CALIBRATION, Cahillane:2017}.
The maximum 1-$\sigma$ calibration uncertainties for strain data used in this analysis are 5\% in amplitude and 3$^{\circ}$ in phase over the frequency range 20--1024\,Hz.

The Advanced Virgo detector was, at the time of the event, in observation mode with a horizon distance for signals comparable to \EventName{} of $60{-}70\,$Mpc.
This was however during an early commissioning phase with still limited sensitivity, therefore Virgo data are not included in the analyses presented here.

\section{Search for binary merger signals}
\label{s:observation}

\begin{figure}
\includegraphics{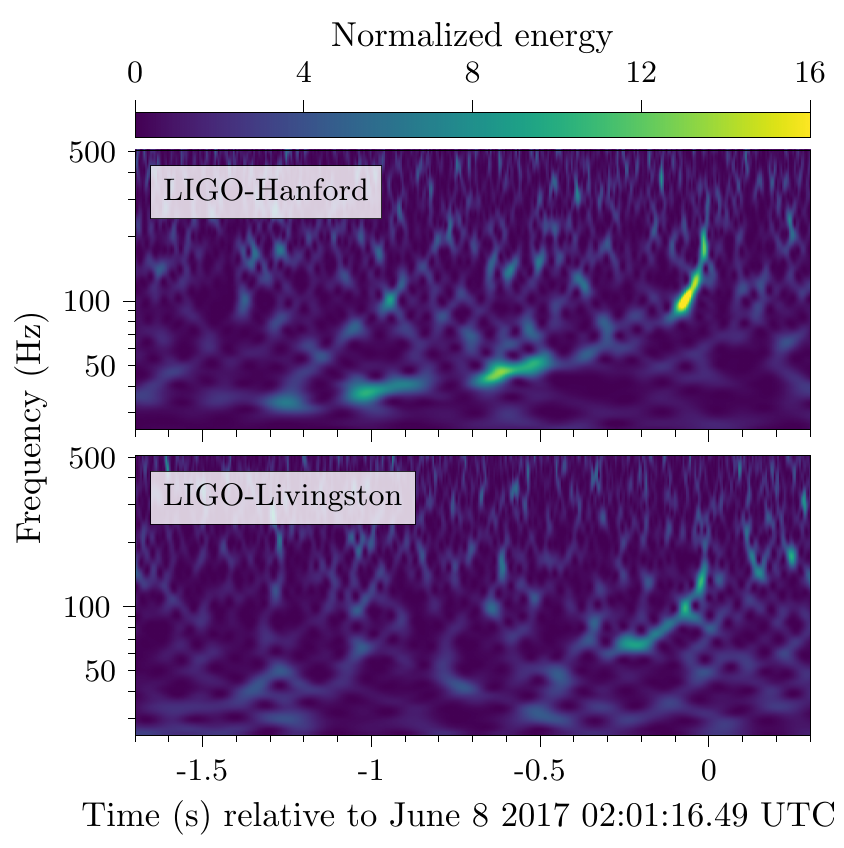}
\caption{Time-frequency power maps of \LIGO strain data at the time of \EventName{}.  The characteristic upwards-chirping morphology of a binary inspiral driven by \GW emission is visible in both detectors, with a higher signal amplitude in \ac{LHO}. 
This figure, and all others in this Letter, were produced from noise-subtracted data (Section~\ref{s:parameters}).
\label{fig:qscans}
}
\end{figure}

\subsection{Low latency identification of a candidate event}
\EventName{} was first identified as a loud (\ac{SNR} ${\sim}9$) 
event in \ac{LLO} data, via visual inspection of single-detector events from a low-latency compact binary matched filter  (`template') analysis~\citep{Usman:2015kfa,Nitz:2017,pycbc-software}.  Such events are displayed automatically to diagnose changes in detector operation and in populations of non-Gaussian transient noise artifacts (glitches)~\citep{GW150914-DETCHAR}.  Low-latency templated searches~\citep{Cannon:2015gha,Messick:2016aqy,Adams:2015ulm,Nitz:2017} did not detect the event with high significance because \ac{LHO} data were not analysed automatically. 
The morphology of the \ac{LLO} event is consistent with a compact binary merger signal, as shown in Figure~\ref{fig:qscans} (lower panel), but a noise origin could not be ruled out using \ac{LLO} data alone. 

Consequently, \ac{LHO} data were investigated and were determined to be stable at frequencies above \SpecialAnalysisFLower{} (see Appendix~\ref{s:app_a2l}). 
A segment of \ac{LHO} data around the event time was then searched with a filter starting frequency of \SpecialAnalysisFLower{}, using templates approximating the waveforms from compact binary systems with component spins aligned with the orbital angular momentum \citep{Bohe:2016gbl, Purrer:2015tud}.  The fraction of \ac{SNR} expected to be lost due to imposing the \SpecialAnalysisFLower{} cutoff, as compared to the lower starting frequencies typically used in O2 data~\citep{DalCanton:2017}, is ${\sim} 1\%$ or less.  An event was found having consistent template binary masses and spins, times of arrival and \acp{SNR} in \ac{LHO} and \ac{LLO}. 
Based on this 2-detector coincident event an alert was issued to electromagnetic observing partners 13.5 hours after the event time, with a sky localization~\citep{Singer:2015ema} covering \EventSkyAreaBayestar\ (90\% credible region). 
GRB Coordinates Network Circulars related to this event are archived at \url{https://gcn.gsfc.nasa.gov/other/G288732.gcn3}.

\subsection{Offline search}
To establish the significance of this coincident event, a period between \SpecialAnalysisStart{} and \SpecialAnalysisEnd{} was identified for analysis during which both \LIGO interferometers were
operating in the same configuration as at the event time.  Times at which commissioning activities at \ac{LHO} produced severe or broad-band disturbances in the strain data were excluded from the analysis.  Standard offline data quality vetoes for known environmental or instrumental artifacts were also applied, resulting overall in \SpecialAnalysisCoincTime{} of coincident \ac{LHO}--\ac{LLO} data searched.

Two matched filter pipelines identified \EventName{}, with a network \ac{SNR} of~\EventNetworkSNR{}.  An candidate event is assigned a ranking statistic value, in each pipeline, that represents its relative likelihood of originating from a \GW signal \emph{vs.}\ from noise.  One pipeline estimates the noise background using time-shifted data~\citep{Usman:2015kfa}, finding a rate of occurrence of noise events ranked higher than \EventName{} of \emph{less} than \EventFARPyCBC{}.  This limit corresponds to the maximum background analysis time available from time shifts separated by $0.1$\,s. 
The other pipeline uses different methods for ranking candidate events and for estimating the background~\citep{Cannon:2015gha,Messick:2016aqy} and assigns the event a false alarm rate of \EventFARGstLALSpecialAnalysis{}. 

A search for transient \GW signals coherent between \ac{LHO} and \ac{LLO} with frequency increasing over time, without using waveform templates~\citep{Klimenko:2015ypf}, also identified \EventName{} with a false alarm rate of \EventFARcWBUnClean; the lower significance is expected as this analysis is typically less sensitive to lower-mass compact binary signals than matched filter searches.

\section{Source Properties}
\label{s:parameters}

\subsection{Binary Parameters}
\label{sec:pe}

The parameters of the \GW source are inferred from a coherent Bayesian analysis~\citep{Veitch:2014wba, GW150914-PARAMESTIM} using noise-subtracted data from the two \LIGO observatories. Noise subtraction is a data processing step that removes several instrumental noise sources from the \GW strain measurements (\cite{GW170814-DETECTION} and references therein), thus increasing the expected \SNR of compact binary signals in \ac{LHO} data by typically $25\%$~\citep{OfflineNoiseSub}.  
The likelihood integration is performed starting at $30$\,Hz in \ac{LHO} and $20$\,Hz in \ac{LLO}, includes marginalization over strain calibration uncertainties~\citep{SplineCalMarg-T1400682}, and uses the noise power spectral densities~\citep{Littenberg:2014oda} at the time of the event.

Two different gravitational-wave signal models calibrated to numerical relativity simulations of general relativistic binary black hole mergers~\citep{Mroue:2013xna, Chu:2015kft, Husa:2015iqa}, building on the breakthrough reported in~\citep{Pretorius:2005gq, Campanelli:2005dd, Baker:2005vv}, are used.
One waveform family models the inspiral-merger-ringdown signal of precessing binary black holes~\citep{Hannam:2013oca}, which includes spin-induced orbital precession through a transformation of the aligned-spin waveform model of~\citep{Husa:2015iqa, Khan:2015jqa}; 
we refer to this model as the \textit{effective precession} model. 
The other waveform model describes binaries with spin angular momenta aligned with the orbital angular momentum~\citep{Bohe:2016gbl, Purrer:2015tud}, henceforth referred to as \textit{non-precessing}.
For their common parameters, both waveform models yield consistent parameter ranges. 

A selection of inferred source parameters for \EventName{} is given in Table~\ref{tab:parameters}; unless otherwise noted, we report median values and symmetric 90\% credible intervals. The quoted parameter uncertainties include statistical and systematic errors from averaging posterior probability samples over the two waveform models. 
As in \cite{GW170104-DETECTION}, our estimates of the mass and spin of the final black hole, the total energy radiated in \acp{GW} as well as the peak luminosity are computed from fits to numerical relativity simulations~\citep{Hofmann:2016yih, Keitel:2016krm, Healy:2016lce, Jimenez-Forteza:2016oae}.
\begin{table}
\caption{Source properties for \EventName{}: given are median values with $90\%$ credible intervals. 
Source-frame masses are quoted; to convert to detector frame, multiply by $(1 + z)$~\citep{Krolak:1987ofj}. 
% TD : 1 reference enough? {Holz:2005df}.
The redshift assumes a flat cosmology with Hubble parameter $H_0 = 67.9~\mathrm{km\,s^{-1}\,Mpc^{-1}}$ and matter
density parameter $\Omega_\mathrm{m} = 0.3065$~\citep{Ade:2015xua}. 
}
\begin{ruledtabular}
\begin{tabular}{l l}
Chirp mass $\mathcal{M}$ & $\MCSCOMPACTJune\,\Msun$ \\
\rule{0pt}{3ex}%
Total mass $M$ & $\MTOTSCOMPACTJune\,\Msun$ \\
\rule{0pt}{3ex}%
Primary black hole mass $m_1$ & $\MONESCOMPACTJune\,\Msun$ \\
\rule{0pt}{3ex}%
Secondary black hole mass $m_2$ & $\MTWOSCOMPACTJune\,\Msun$ \\
\rule{0pt}{3ex}%
Mass ratio $m_2/m_1$ & $\MASSRATIOCOMPACTJune$ \\
\rule{0pt}{3ex}%
Effective inspiral spin parameter $\chi_\mathrm{eff}$ & $\CHIEFFCOMPACTJune$ \\
\rule{0pt}{3ex}%
%Final black hole mass $M_\mathrm{f}$ & $\MFINALSCOMPACTJune\,\Msun$ \\
%\rule{0pt}{3ex}%
%Use when available: 
Final black hole mass $M_\mathrm{f}$ & $\MFINALSavgCOMPACTJune\,\Msun$ \\
\rule{0pt}{3ex}%
Final black hole spin $a_\mathrm{f}$ & $\SPINFINALinplaneCOMPACTJune$ \\
\rule{0pt}{3ex}%
%Radiated energy $E_\mathrm{rad}$ & $\ERADJune\,\Msun c^{2}$ \\
%\rule{0pt}{3ex}%
%Use 
Radiated energy $E_\mathrm{rad}$ & $\ERADCOMPACTJune\,\Msun c^{2}$ \\
\rule{0pt}{3ex}%
Peak luminosity $\ell_\mathrm{peak} $& $\LPEAKCOMPACTJune\,\mathrm{erg\,s^{-1}}$ \\
\rule{0pt}{3ex}%
%Final black hole spin $a_\mathrm{f}$ & $\SPINFINALCOMPACTJune$ \\
%\rule{0pt}{3ex}%
%Use when available: 
Luminosity distance $D_\mathrm{L}$ & $\DISTANCECOMPACTJune\,\mathrm{Mpc}$ \\
\rule{0pt}{3ex}%
Source redshift $z$ & $\REDSHIFTCOMPACTJune$ \\
\end{tabular}
\end{ruledtabular}
\label{tab:parameters}
\end{table}
\begin{figure} 
\includegraphics{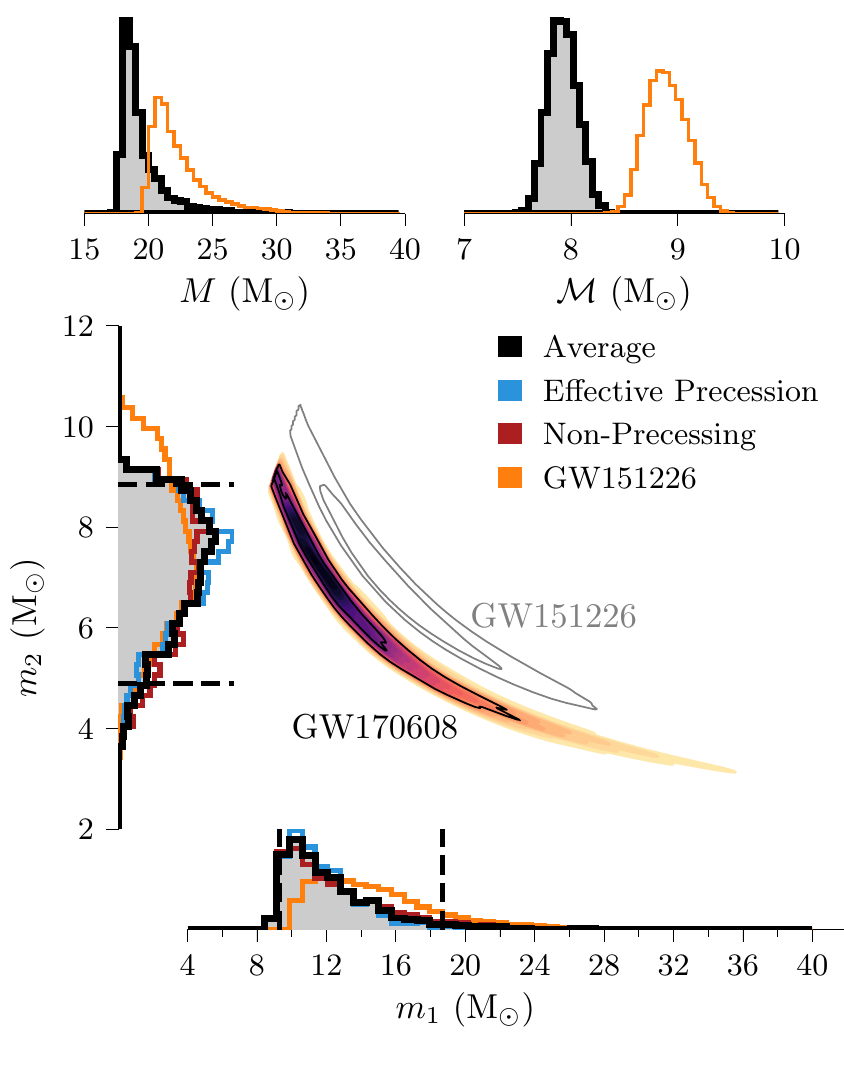} 
\caption{Posterior probability densities for binary component masses ($m_1$, $m_2$), total mass ($M$), and chirp mass ($\mathcal{M}$) in the source frame. One-dimensional component mass distributions include posteriors for the effective precession (blue) and the non-precessing (red) waveform model, as well as their average (black). The dashed lines demarcate the $90\%$ credible intervals for the average posterior. The two-dimensional plot shows contours of the $50\%$ and $90\%$ credible regions overlaid on a color-coded posterior density function. For comparison, we show both one- and two-dimensional distributions of averaged component mass posterior samples for GW151226 (orange)~\citep{GW151226-DETECTION}. In the top panel, we further compare \EventName{} and GW151226's source-frame total mass (left) and source-frame chirp mass (right). All other known binary black holes lie at higher chirp masses than \EventName{} and GW151226.}
\label{fig:masses}
\end{figure}

The posterior probability distributions for the source-frame mass parameters of \EventName{} 
are shown in Figure~\ref{fig:masses}, together with those for GW151226~\citep{GW151226-DETECTION}. 
The initial binary of \EventName{} consisted of two compact objects with source-frame component masses $m_1= \MONESCOMPACTJune\,\Msun$ and $m_2= \MTWOSCOMPACTJune\,\Msun$, with the mass ratio loosely constrained to $m_2/m_1 = \MASSRATIOCOMPACTJune$. 
Since neutron stars are expected to have masses below ${\sim} 4 M_{\odot}$~\citep{Lattimer:2015nhk}, both objects are most likely black holes.  
Notably, we find this binary black hole system to be the least massive yet observed through gravitational waves. 
The next lightest, GW151226~\citep{GW151226-DETECTION}, 
has a chirp mass $\mathcal{M} = \MCSCOMPACTBoxing$ and a total mass $M = \MTOTSCOMPACTBoxing$, compared with values of $\mathcal{M}=\MCSCOMPACTJune\,\Msun$ and $M=\MTOTSCOMPACTJune\, \Msun$ for \EventName{}. 
The probability that \EventName{}'s total mass is smaller than GW151226's is $0.89$.

While the chirp mass is tightly constrained, spins have a more subtle effect on the \GW signal. 
The effective inspiral spin $\chi_\mathrm{eff}$, a mass-weighted combination of the spin components (anti-)aligned with the orbital angular momentum~\citep{Racine:2008qv, Ajith:2009bn}, predominantly affects the inspiral rate of the binary but also influences the merger. We infer that $\chi_\mathrm{eff}=\CHIEFFCOMPACTJune$ disfavoring large, anti-aligned spins on both black holes. 

An independent parameter estimation method comparing \LIGO strain data to hybridized numerical relativity simulations of binary black hole systems with non-precessing spins~\citep{GW150914-NRCOMP} yields estimates of component masses and $\chi_\mathrm{eff}$ consistent with our model-waveform analysis.

\begin{figure}
\includegraphics{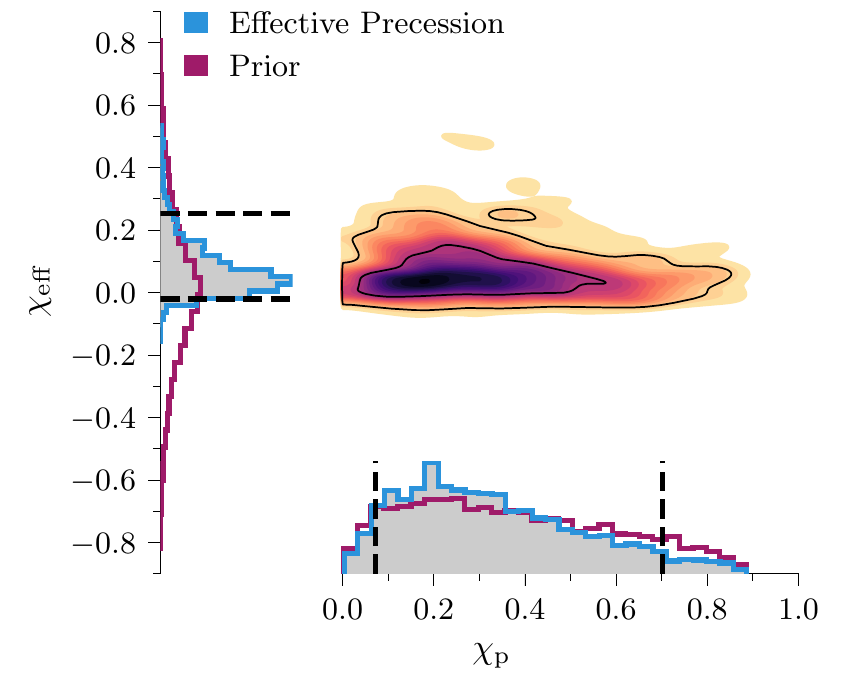}
\includegraphics{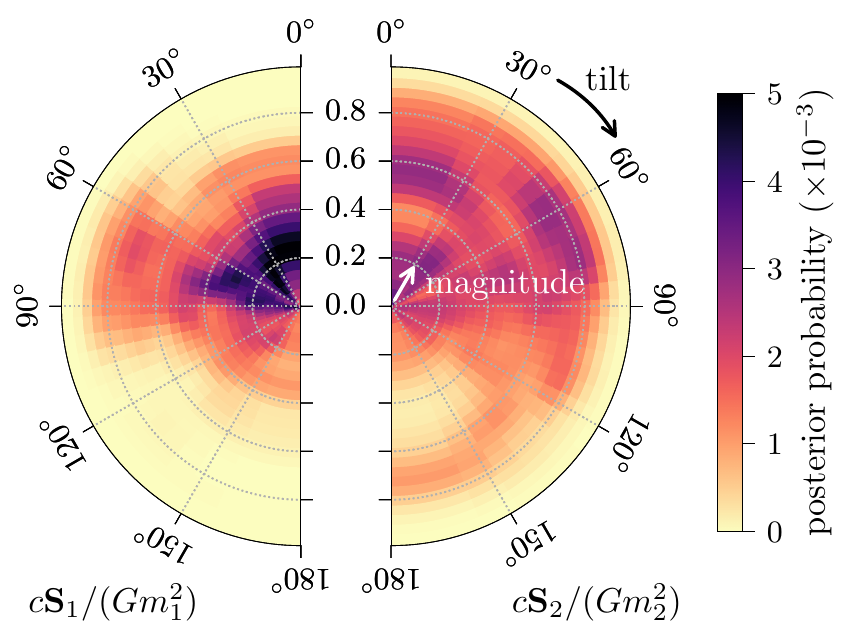}
\caption{\textit{Top panel}: Marginalized one-dimensional posterior density functions for the spin parameters $\chi_p$ and $\chi_\mathrm{eff}$ (blue) in comparison to their prior distributions (pink) as obtained from the effective-precession model. 
The dashed lines indicate the 90\% credible interval. The two-dimensional plot shows the 50\% and 90\% credible regions plotted over the posterior density function.
\textit{Bottom panel}: Posterior probabilities for the dimensionless component spins 
$\mathbf{\chi}_i$ with $i=1,2$ relative to the Newtonian orbital angular momentum $\hat{\mathbf{L}}$, i.e.\ the normal of the orbital plane. The tilt angles are $0^\circ$ for spins parallel to $\hat{\mathbf{L}}$, and $180^\circ$ for spins anti-parallel to $\hat{\mathbf{L}}$. The posterior density functions are marginalized over the azimuthal angles. Each pixel has a prior probability of $\sim1.8 \times 10^{-3}$; they are spaced linearly in spin magnitudes and the cosine of the tilt angles. 
\label{fig:spins}}
\end{figure}

Spin components orthogonal to the orbital angular momentum are the source of precession \citep{Apostolatos:1994mx, Kidder:1995zr}, and may be parameterized by a single effective precession spin $\chi_p$~\citep{Schmidt:2014iyl}. 
For precessing binaries, component spin orientations evolve over time; we report results evolved to a reference \GW frequency of $20$\,Hz. 
The spin prior assumed in this analysis is uniform in dimensionless spin magnitudes $\mathbf{\chi}_i \equiv c|\mathbf{S}|_i/(Gm_i^2)$ with $i=1,2$ between $0$ and $0.89$, and isotropic in their orientation; this prior on component spins maps to priors for the effective parameters $\chi_\mathrm{eff}$ and $\chi_p$. 
The top panel of Figure~\ref{fig:spins} shows the prior and posterior probability distributions of $\chi_\mathrm{eff}$ and $\chi_p$ obtained for the effective-precession waveform model. While we gain some information about $\chi_\mathrm{eff}$, the $\chi_p$ posterior is dominated by its prior, as for previous GW events~\citep{GW150914-DETECTION, GW151226-DETECTION, GW170104-DETECTION}, indicating that we cannot draw any strong conclusion on the size of spin components in the orbital plane~\citep{Vitale:2017}. 
The inferred component spin magnitudes and orientations are shown in the bottom panel of Figure~\ref{fig:spins}. We find the dimensionless spin magnitude of the primary black hole, $\chi_1$, to be less than $0.75$ ($90\%$ credible limit); this limit is robust to extending the prior range of spin magnitudes and to using different waveform models.

The measurability of precession depends on the intrinsic source properties as well as the angle of the binary orbital angular momentum to the line of sight (i.e.\ inclination). 
The inclination of \EventName{}'s orbit is likely close to either $0^\circ$ or $180^\circ$, due to a selection effect: the distance inside which a given binary merger would be detectable at a fixed \ac{SNR} threshold is largest for these inclination values \citep{Schutz:2011tw}.  For such values, the waveform carries little information on precession.

The distance of \EventName{} is extracted from the observed signal amplitude given the binary's inclination~\citep{GW150914-PARAMESTIM}. With the network of two nearly co-aligned \LIGO detectors, the uncertainty on inclination translates into a large distance uncertainty: we infer a luminosity distance of $D_L = \DISTANCECOMPACTJune~\mathrm{Mpc}$, corresponding to a redshift of $z=\REDSHIFTCOMPACTJune$ assuming a flat $\Lambda\mathrm{CDM}$ cosmology~\citep{Ade:2015xua}. 

\EventName{} is localized to a sky area of ${\sim}\PESKYNINTYJune$ in the Northern hemisphere (90\% credible region), determined largely by the signal's measured arrival time at \ac{LLO} ${\sim}\EventHLAbsTimeDelay$ later than at \ac{LHO}. 
This reduction in area relative to the low-latency map is partly attributable to the use of noise-subtracted data with offline calibration~\citep{GW170814-DETECTION}.

\subsection{Consistency with General Relativity}
\label{sec:tgr}
To test whether \EventName{} is consistent with the predictions of \GR, we consider possible deviations of coefficients describing the binary inspiral part of the signal waveform from the values expected in \GR, as was done for previous detections~\citep{GW150914-TESTOFGR,O1:BBH,GW170104-DETECTION}. 
Tests involving parameters describing the merger and ringdown do not yield informative results, since the merger happens at relatively high frequency where the \LIGO detectors are less sensitive.
As in \cite{GW170814-DETECTION}, we also allow a sub-leading phase contribution at effective $-1$PN order, i.e.\ with a frequency dependence of $f^{-7/3}$, which is absent in \GR.  The \GR predicted value is contained within the $90\%$ credible interval of the posterior distribution for all parameters tested.

Assuming that gravitons are dispersed in vacuum like massive particles, we also obtained an upper bound on the mass of the graviton comparable to the constraints previously obtained \citep{GW150914-DETECTION, GW150914-TESTOFGR, GW170104-DETECTION}. Possible violations of local Lorentz invariance, manifested via modifications to the \GW dispersion relation, were investigated~\citep{GW170104-DETECTION}, again finding upper bounds comparable to previous results.

\section{Astrophysical Implications}
\label{s:astro}

The low mass of \EventName{}'s source binary, in comparison to other binary black hole systems observed by \LIGO and Virgo, 
has potential implications for the binary's progenitor environment. High-metallicity progenitors are expected to experience substantial mass loss through strong stellar winds~\citep{Spera:2015vkd}, implying that high-mass black hole binaries are not formed in such environments. \EventName{}'s low mass suggests formation in a higher metallicity environment; however, formation at lower metallicity with comparatively lower mass progenitors is not excluded. Further discussion of the relationship between black hole masses and metallicity can be found in~\cite{GW150914-ASTRO}.

At this lower boundary of the observed distribution of binary black hole masses, we can compare component black holes with those found in X-ray binaries. X-ray binary systems contain either a black hole or neutron star which accretes matter from a companion donor star. \ac{LMXBs} are X-ray binaries with a low-mass donor star which transfer mass through Roche lobe overflow~\citep{Charles:2003rj}. The inferred component masses of \EventName{} are consistent with dynamically-measured masses of black holes found in \ac{LMXBs}, typically less than 10$\Msun$~\citep{Ozel:2010su, Farr:2010tu, Corral-Santana:2015fud}. 

Black holes in \ac{LMXBs} are believed to form with near-zero spin and acquire spin as a byproduct of mass accretion~\citep{Fragos:2014cva}. For \EventName{}, we infer an effective spin probability distribution that is concentrated around zero with the 90\% credible interval extending to small positive spin. Thus we can exclude highly negative anti-parallel spin components, while remaining broadly consistent with expected LMXB spins~\citep{Miller:2014aaa, Fragos:2014cva}. 

Binary black holes may form through many different channels, including, but not limited to, dynamical interaction~\citep{Rodriguez:2016avt, OLeary:2016ayz, Mapelli:2016vca} and isolated binary evolution~\citep{Belczynski:2016obo, Eldridge:2016ymr, Lipunov:2016cml, Stevenson:2017tfq}. While the inferred masses and tilt measurements of \EventName{} are not sufficiently constrained to favor a formation channel, future measurements of binary black hole systems may hint at the formation histories of such systems (see~\cite{GW170104-DETECTION} and references therein). It may be possible to determine the relative proportion of binaries originating in each canonical formation channel following $\mathcal{O}(100)$ binary black hole detections~\citep{Vitale:2015tea, Stevenson:2017dlk, Zevin:2017, Talbot:2017yur, BFarr:2017, Farr:2017}.

The detection of \EventName{} is consistent with the merger populations considered in \cite{GW150914-RATES,O1:BBH} for which a rate of \rateintrbracket{} was estimated in \cite{GW170104-DETECTION}.

\section{Outlook}
\label{s:summary}

\LIGO's detection of \EventName{} extends the range of known stellar-mass binary black hole systems at the low-mass boundary, and hints at connections with other known astrophysical systems containing black holes. 
The \ac{O2} run ended on August 25th, 2017;  
a full catalog of binary merger gravitational-wave events for this run is in preparation, including candidate signals with lower significance and systems other than stellar-mass black hole binaries \citep{GW170817-DETECTION}.  Estimates of the merger rate and mass distribution for the emerging compact binary population will also be updated. 

With expected increases in detector sensitivity in the third advanced detector network observing run, projected for late 2018 \citep{Aasi:2013wya}, detection of black hole binaries will be a routine occurrence; studying this population will eventually answer many questions about these systems' origins and evolution.

\acknowledgments 
The authors gratefully acknowledge the support of the United States
National Science Foundation (NSF) for the construction and operation of the
LIGO Laboratory and Advanced LIGO as well as the Science and Technology Facilities Council (STFC) of the
United Kingdom, the Max-Planck-Society (MPS), and the State of
Niedersachsen/Germany for support of the construction of Advanced LIGO 
and construction and operation of the GEO600 detector. 
Additional support for Advanced LIGO was provided by the Australian Research Council.
The authors gratefully acknowledge the Italian Istituto Nazionale di Fisica Nucleare (INFN),  
the French Centre National de la Recherche Scientifique (CNRS) and
the Foundation for Fundamental Research on Matter supported by the Netherlands Organisation for Scientific Research, 
for the construction and operation of the Virgo detector
and the creation and support  of the EGO consortium. 
The authors also gratefully acknowledge research support from these agencies as well as by 
the Council of Scientific and Industrial Research of India, 
the Department of Science and Technology, India,
the Science \& Engineering Research Board (SERB), India,
the Ministry of Human Resource Development, India,
the Spanish  Agencia Estatal de Investigaci\'on,
the  Vicepresid\`encia i Conselleria d'Innovaci\'o, Recerca i Turisme and the Conselleria d'Educaci\'o i Universitat del Govern de les Illes Balears,
the Conselleria d'Educaci\'o, Investigaci\'o, Cultura i Esport de la Generalitat Valenciana,
the National Science Centre of Poland,
the Swiss National Science Foundation (SNSF),
the Russian Foundation for Basic Research, 
the Russian Science Foundation,
the European Commission,
the European Regional Development Funds (ERDF),
the Royal Society, 
the Scottish Funding Council, 
the Scottish Universities Physics Alliance, 
the Hungarian Scientific Research Fund (OTKA),
the Lyon Institute of Origins (LIO),
the National Research, Development and Innovation Office Hungary (NKFI), 
the National Research Foundation of Korea,
Industry Canada and the Province of Ontario through the Ministry of Economic Development and Innovation, 
the Natural Science and Engineering Research Council Canada,
the Canadian Institute for Advanced Research,
the Brazilian Ministry of Science, Technology, Innovations, and Communications,
the International Center for Theoretical Physics South American Institute for Fundamental Research (ICTP-SAIFR), 
the Research Grants Council of Hong Kong,
the National Natural Science Foundation of China (NSFC),
the Leverhulme Trust, 
the Research Corporation, 
the Ministry of Science and Technology (MOST), Taiwan
and
the Kavli Foundation.
The authors gratefully acknowledge the support of the NSF, STFC, MPS, INFN, CNRS and the
State of Niedersachsen/Germany for provision of computational resources.

\appendix

\section{Angular coupling minimization}
\label{s:app_a2l}

\EventName{} was observed during a routine instrumental procedure at \ac{LHO} that minimises the coupling of angular control of the test masses to noise in the \GW strain measurement. 
To maintain resonant power in the arms, the pitch and yaw angular degrees of freedom of the four suspended cavity test masses at each detector~\citep{GW150914-DETECTORS} must be controlled.  This is achieved by actuating on the second stage of the LIGO quadruple suspensions.
A feed-forward control is employed in order to leave the beam position of the main laser on the test mass unchanged while this actuation is applied. 
However, if this position differs from the actuation point, the angular control can affect the differential arm length, thus introducing additional noise in the strain measurement~\citep{A2LBeamPos}.
As the beam position can drift over periods of hours or days, the angular feed-forward control must be periodically adjusted in order to minimize the coupling to strain.

During this procedure, high amplitude pitch and yaw excitations are applied to the test masses 
via actuation of the suspensions. 
Each of the 8 angular degrees of freedom is excited at a distinct frequency;
the resulting length signals are observed via demodulation at each excitation frequency, revealing how strongly the corresponding degree of freedom couples to differential arm length.  The feed-forward gain settings are stepped at intervals of approximately $45$\,s and the global minimum of angular control coupling to strain is determined from the resulting measurements. 
\begin{figure}[bp]
  \includegraphics{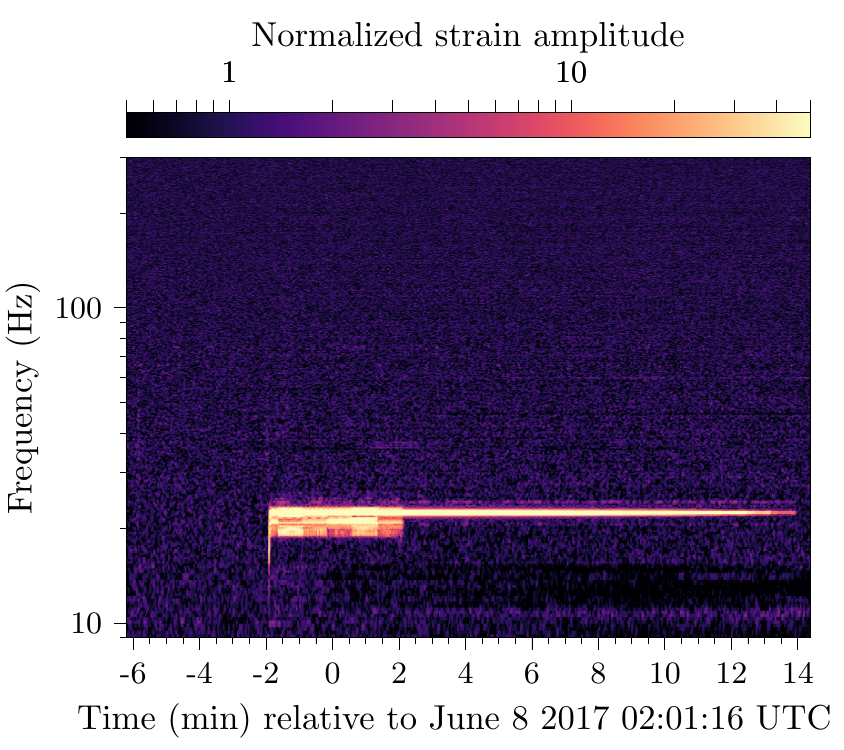}
  \includegraphics{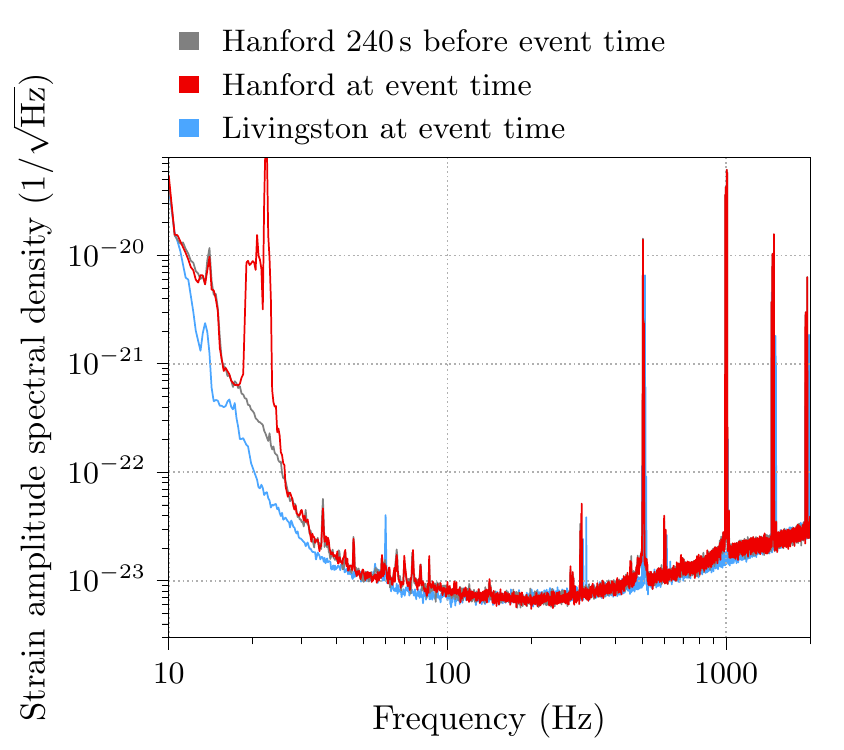}
  \caption{Left: Spectrogram of strain data from \ac{LHO} around the time of \EventName{}.  
           This plot shows variations in the noise spectrum of the detector over periods
           on the scale of minutes; unlike Figure 1, it is not designed to show short-duration
           transient events.  The strain amplitude is normalized to the interval between
           $-6$ and $-2$ minutes relative to the event time.  See Appendix~\ref{s:app_a2l} for 
           discussion of the feature around $20$\,Hz due to an angular control procedure. 
           Right: Amplitude spectral density of strain data at both LIGO observatories for 
           240\,s around the event time, ($-2,2$) minutes on left panel, and for data 
           before the start of the angular coupling minimization at \ac{LHO}, ($-6,-2$) minutes.
           Excess noise is clearly visible around $20$\,Hz but data above
           \SpecialAnalysisFLower{} are unaffected.  
 \label{fig:a2l_spec}}
\end{figure}
The frequencies of angular excitations are equally spaced between ${\sim} 19$\,Hz and ${\sim} 23$\,Hz, generating excess power in the differential arm motion, and thus in the measured strain around these frequencies.
This procedure covers from ${\sim} 2$ minutes before to ${\sim} 14$ minutes after \EventName{}, shown in Figure~\ref{fig:a2l_spec} (left).
During the period from $-2$ to $2$ minutes substantial excess noise is visible at frequencies around 20\,Hz.
To characterize this noise we show amplitude spectral densities derived from $240$\,s of data both before the onset of the angular excitations and during the excitations around the event time in Figure~\ref{fig:a2l_spec} (right).
No effect on the spectrum is visible above \SpecialAnalysisFLower{}.

During the procedure, angular control gain settings are stepped abruptly; inspection of all such transition times shows no evidence for transient excess noise in the strain data outside the $19$--$23$\,Hz excitation band. 
The closest transition to the event time was $10$\,s before the binary merger, thus any transient noise associated with this transition could not have affected the matched-filter output at the event time (template waveforms for \EventName{}-like signals have a duration between $2$ and $3$\,s.) 
Furthermore, events from a single-detector matched-filter search covering other periods at \ac{LHO} when this procedure was performed shows no anomalous features compared to other times.  Thus, we find no evidence that the angular coupling minimization affected the recorded strain data at \ac{LHO} around the event time at frequencies above \SpecialAnalysisFLower{}.

\end{document}